\documentclass[twocolumn,amssymb,nobibnotes,superscriptaddress,aps,pra]{revtex4}

\usepackage{enumerate,amsmath}
\usepackage{graphicx}
\usepackage[dvips]{color} 

\begin{document}

\title{
Measurement-Based Quantum Computation on Symmetry Breaking Thermal States
}%

\author{Keisuke Fujii}
\affiliation{
Graduate School of Engineering Science, Osaka University,
Toyonaka, Osaka 560-8531, Japan}

\author{Yoshifumi Nakata}
\affiliation{
Department of Physics, Graduate School of Science, University of Tokyo, Tokyo 113-0033, Japan}

\author{Masayuki Ohzeki}
\affiliation{
Department of Systems Science, Graduate School of Informatics, Kyoto University, Yoshida-Honmachi, Sakyo-ku, Kyoto 606-8501, Japan}

\author{Mio Murao}
\affiliation{
Department of Physics, Graduate School of Science, University of Tokyo, Tokyo 113-0033, Japan}
\affiliation{
Institute for Nano Quantum Information Electronics, University of Tokyo, Tokyo 153-8505, Japan}

\date{\today}
\begin{abstract}
We consider measurement-based quantum computation (MBQC) on thermal states of the 
interacting cluster Hamiltonian containing interactions between the cluster stabilizers that undergoes thermal phase transitions.
We show that the long-range order of the symmetry breaking thermal states below a critical temperature
drastically enhance the robustness of MBQC against thermal excitations. 
Specifically, we show the enhancement in two-dimensional cases and prove that MBQC is topologically
protected  below the critical temperature in three-dimensional cases.
The interacting cluster Hamiltonian allows us to perform
MBQC even at a temperature an order of magnitude higher 
than that of the free cluster Hamiltonian.
\end{abstract}


\maketitle
 {\it Introduction.---}
Measurement-based quantum computation (MBQC) is a paradigm
for quantum computation, where a many-body entangled state
is prepared as a universal resource state, 
and quantum computation
can be executed by adaptive single-qubit measurements on it \cite{OWC}.
This paradigm
provides a good clue to understand requirements on a system as a resource
for universal quantum computation,
making a bridge between quantum information science
and many-body physics.
A central issue in this approach is 
to specify a many-body system
whose ground or low-temperature thermal state
can serve as a universal resource for MBQC.

Ground states of several Hamiltonians such as
the cluster Hamiltonians \cite{RBH,VC,BR}
and valence-bond solid Hamiltonians \cite{Brennen08,Chen09,Cai10,Wei11,Miyake11},
have been found to be universal.
At finite temperature, 
thermal states of several Hamiltonians
have been shown to be useful as universal resources by protecting 
quantum information from errors originating from the thermal excitation by using quantum error correction
\cite{Barrett09,Li11,FM12}.
However, these Hamiltonians do not undergo any physical (thermal nor quantum) phase transitions, although they exhibit a transition in computational capability of MBQC by varying temperature.
Thus we address a question
whether it is possible to enhance the robustness of MBQC
against thermal excitations by introducing
a Hamiltonian that undergoes a phase transition. 
This can strengthen the connection between quantum information science and many-body physics and can provide an approach to understand
the robustness of MBQC in terms of many-body physics. 

In this letter, we show that robustness of MBQC on thermal states 
can be enhanced drastically by introducing interactions between the cluster stabilizers.
The proposed Hamiltonian --{\it interacting} cluster Hamiltonian (iCH)-- is transformed into a ferromagnetic Ising Hamiltonian by unitary transformations. Hence, it undergoes a phase transition on two or higher dimensional lattices, leading to the symmetry breaking of the thermal states.
By virtue of the ferromagnetic-type long-range order of such symmetry breaking states, MBQC becomes robust below the 
critical temperature.
We first demonstrate this on a two-dimensional (2D) lattice
and show that the fidelity of MBQC can be drastically improved below the critical temperature due to the long-range order, although 
it is not sufficiently large at the temperature just below the critical temperature.
We further investigate topologically protected MBQC on a three-dimensional (3D) lattice \cite{Kitaev,Dennis,Wang,Ohno,RBH,RaussendorfA,RaussendorfB,RaussendorfC} in order to achieve quantum computation of arbitrary accuracy at any temperature below the critical temperature.
We show that the threshold value for the topologically protected MBQC is exactly equal to the critical temperature.
Compared to the previous Hamiltonian without the interactions between the cluster stabilizers~\cite{RBH},
the temperature required for topologically protected
MBQC is relaxed by more than one order of magnitude.

\begin{figure}
\begin{center}
\includegraphics[width=85mm]{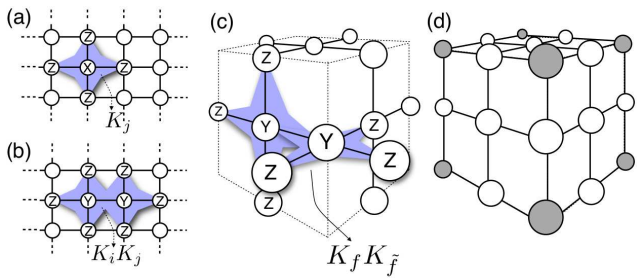}
\caption{(a) The fCH on the square lattice. (b) The iCH on the square lattice.
 (b) The interacting cluster Hamiltonian on the RHG lattice.
 (c) The 3D cluster state on the cubic lattice. (d) The gray-shaded qubits are measured
 in the $Z$ basis to obtain the cluster state on the RHG lattice.}
\label{fig1}
\end{center}
\end{figure}

{\it Cluster Hamiltonian.---}
The {\it cluster stabilizer} on a lattice $\mathcal{T}$ is given by 
$K_i = X_i \bigotimes _{j\in V_i} Z_j$ for each site $i$, where
$A_i$ ($A=X,Y,Z$) are the Pauli operators on the $i$th qubit, and $V_i$ denotes the set of the vertices
that are adjacent to the site $i$ in the lattice $\mathcal{T}$ \cite{OWC}.
The cluster state on the lattice $\mathcal{T}$, $|\Psi _{\mathcal{T}}\rangle$, is defined by the simultaneous eigenstate
of all cluster stabilizers $K_i$ with eigenvalue $+1$.
The cluster Hamiltonian is defined by using the cluster stabilizers
as $H_{\rm fc} = -J \sum _i K_i$ \cite{RBH} (see Fig. \ref{fig1}(a)),
where $J$ is a coupling constant.
It is obvious that this Hamiltonian has
the cluster state $|\Psi _{\mathcal{T}}\rangle$ as its ground state.
By using a unitary transformation $U_{\mathcal{T}}$,
the products of controlled-$Z$ gates on
all bonds of the lattice $\mathcal{T}$,
the Hamiltonian can be transformed into an interaction-free Hamiltonian 
$U_{\mathcal{T}} H_{\rm fc} U_{\mathcal{T}}= -J \sum _i  X_i $.
We call this Hamiltonian a {\it free cluster} Hamiltonian (fCH) hereafter.
It should be emphasized that this system does not exhibit any phase 
transition since the thermal state is equivalent to that for the interaction-free Hamiltonian.

The thermal state 
can be calculated as
$
\left(\prod _{i} \mathcal{E}_i \right) | \Psi \rangle \langle \Psi |,
$
where $\mathcal{E}_{i} \rho=( \rho + e^{ - 2 \beta J} Z_i \rho Z_i)/(1+e^{- 2\beta J})$
\cite{RBH},
which can be regarded as a cluster state with
an independent $Z$ error on each qubit with probability 
$p_{\beta J}= e^{-2\beta J}/(1+e^{- 2\beta J})$.
It has been known that
it is possible to perform MBQC
in a topologically protected way
with a 3D cluster state
on the so-called Raussendorf-Harrington-Goyal (RHG) lattice \cite{RaussendorfA,RaussendorfB,RaussendorfC}.
The threshold values 
have been numerically calculated to be $p_{\beta J}= 2.9$--$3.3\%$ 
(i.e.  $T_c=1/(\beta _c J) = 0.57$--$0.59$)
depending on the decoding algorithms \cite{Wang,Ohno}.
These threshold values
correspond to the critical concentrations of antiferromagnetic interaction
at zero temperature and at the multicritical point (on the Nishimori line \cite{Nishimori})
of the random plaquette $Z_2$ gauge model (RPGM) \cite{Dennis,Wang,Ohno}.

Next we introduce an {\it interacting cluster} Hamiltonian (iCH),
$H_{\rm ic} = -J \sum _{\langle ij \rangle} K_i K_j$
where the summation runs over
all bonds $\langle i j \rangle$ of the lattice $\mathcal{T}$, and thus each cluster stabilizer interacts
with its nearest neighbors
as shown in Fig.~\ref{fig1} (b) for the case of the square lattice.
This indicates that the iCH generally contains mutually dependent stabilizer operators.
The iCH can be transformed 
to the ferromagnetic Ising Hamiltonian on the lattice $\mathcal{T}$ by $U_{\mathcal{T}}$ as 
$U_{\mathcal{T}} H_{\rm ic} U_{\mathcal{T}} ^{\dag}
= -J \sum _{\langle ij \rangle} X_i X_j$, which is denoted by $H_{\rm Ising}$. 
If the geometrical structure of the lattice
is chosen properly, for example, two or higher dimensional lattices, phase transitions happen.
Although each eigenstate of the iCH, $|\Psi \rangle$, is degenerate with $ \prod _{i \in \mathcal{T}} Z_i |\Psi \rangle$ due to the symmetry of the iCH, the symmetry breaking takes place below the critical temperature.
Hence, the population of either $|\Psi \rangle$ or $ \prod _{i \in \mathcal{T}} Z_i |\Psi \rangle$ becomes much larger than that of the another, which results in a symmetry breaking thermal state exhibiting a long-range order.
We can identify which of the symmetry breaking states
is realized by measuring a cluster stabilizer.
MBQC on such symmetry breaking thermal states
would exploit robustness of long-range order.

{\it MBQC on the square lattice iCH.---}
Let us first consider MBQC in the iCH on a square lattice where a periodic boundary condition is assumed.
Since the iCH is unitarily equivalent to the Ising Hamiltonian, 
the system undergoes a phase transition at the critical temperature 
$T_c/J= 2/\ln[1+\sqrt{2}]$~\cite{Onsager}. 
To check whether the symmetry breaking thermal state leads to the robustness of MBQC,
we consider to perform the identity and Hadamard gates.

For performing the identity and Hadamard gates, we remove unnecessary qubits on the square lattice by measuring them in the $Z$ basis and obtain a cluster state on a line~\cite{OWC}. 
To transfer a state of the first qubit on the line to the $l$-th qubit, we measure all qubits up to the $(l-1)$-th qubit in the $X$ basis and obtain the measurement outcomes $m= \{m_i\} \in \{0,1\}^{\times (l-1)}$. Then we apply the Pauli bi-product to the $l$-th qubit, $B_m=X^{r_m^X} Z^{r_m^Z}$ where $r_m^X = \sum_{i=1}^{\lceil (l-1)/2 \rceil} m_{2i}$ and $r_m^Z = \sum_{i=1}^{\lceil l/2 \rceil} m_{2i-1}$. The summation is taken by modulo $2$ and $\lceil a \rceil$ is the smallest integer larger or equal to $a$.
When the number of the qubits that are measured in the $X$ basis is even (odd), 
the identity (Hadamard) gate is implemented~\cite{OWC}.

\begin{figure}
\begin{center}
\includegraphics[width=42mm]{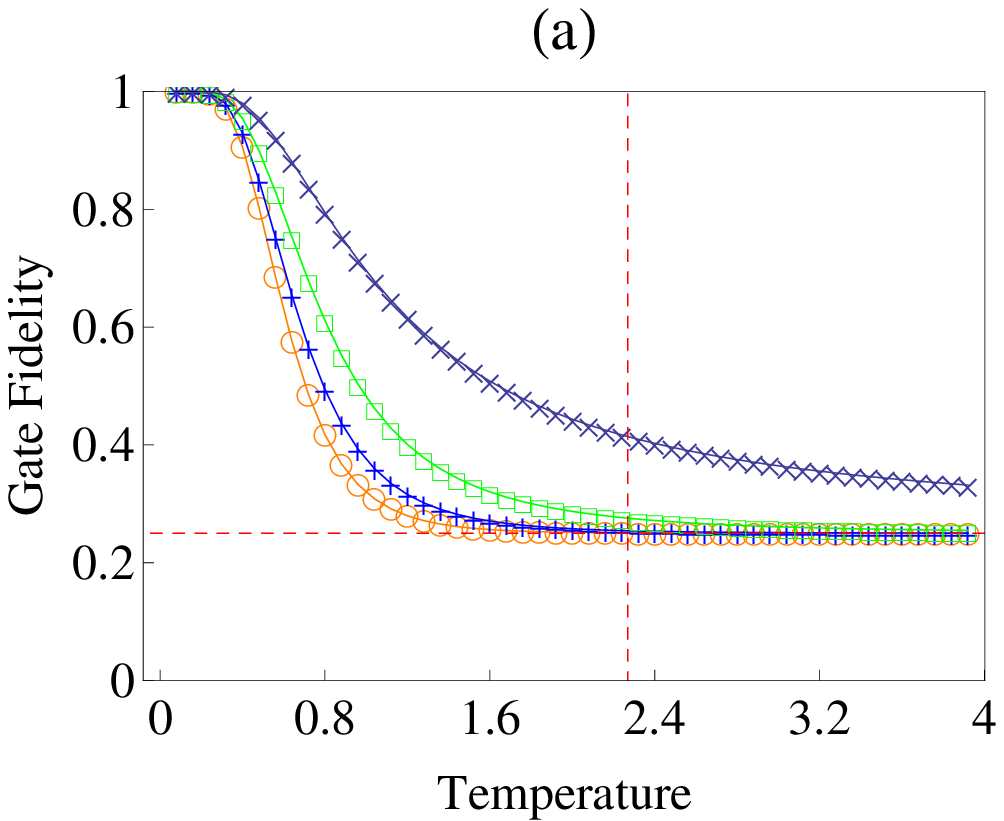}
\includegraphics[width=42mm]{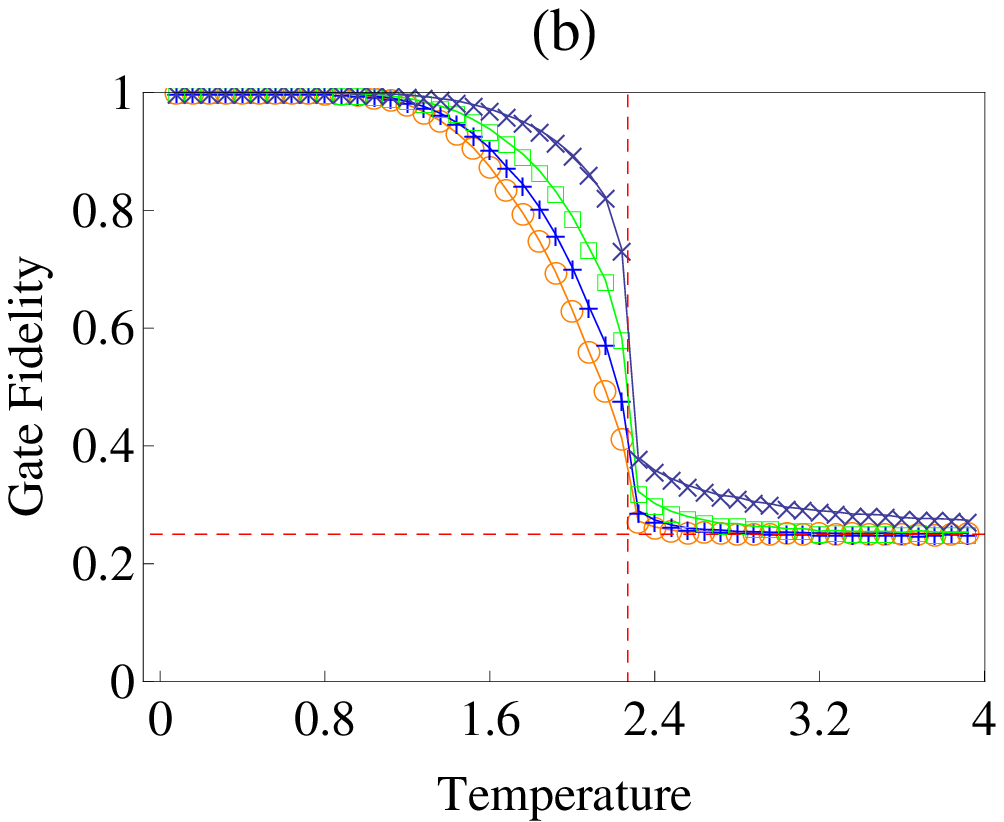}
\caption{The gate fidelities of the identity gates for various distance $l$. Fig. (a) shows the gate fidelity for fCH,
$F (2l)$ for $l=2$ ($\times$), $l=4$ ($\Box$), $l=6$ ($+$) and $l=8$ ($\circ$). 
Fig. (b) shows the gate fidelity for iCH,
$F (2l)$ for $l=2$ ($\times$), $l=4$ ($\Box$), $l=6$ ($+$) and $l=8$ ($\circ$). The vertical dashed line shows the critical temperature for the 2D iCH, $T_c/J= 2/\ln[1+\sqrt{2}]$ and 
the horizontal dashed line shows the minimum gate fidelity $F=1/4$.}
\label{Fig:Gate}
\end{center}
\end{figure}

When the MBQC is performed on the thermal states, each measurement outcome includes an error originated from thermal excitations. 
By denoting such an error by $e=\{ e_i \} \in \{0,1\}^{\times (l-1)}$,
the $i$-th measurement outcome containing the error is given by $\tilde{m}_i=m_i + e_i$ (mod $2$) where $m_i$ is the measurement outcome for the ideal cluster state. 
This results in an error $B_e$ of the gate operation through the Pauli bi-product $B_{\tilde{m}}=B_{e}B_{m}$. 
When the error is correlated so that both of $r_e^X$ and $r_e^Z$ are even, $B_{e}=I$, that is, the gate operation does not include any errors.
Thus, the probability of implementing the gates without any errors
is equal to the overlap between the thermal state and the projector onto the subspace where both of $r_e^X$ and the $r_e^Z$ are even.
Such a probability is referred to as a {\it gate fidelity} and is given by
\begin{equation}
F(l) = {\rm Tr} \biggl[ \frac{I + \prod_{i=1}^{\lceil (l-1)/2 \rceil} K_{2i} }{2} \frac{I + \prod_{i=1}^{\lceil l/2 \rceil} K_{2i} }{2} \rho_{\rm th} \biggr],
\end{equation}
where $\rho_{\rm th}$ is a thermal state of a given Hamiltonian $H$, $\rho_{\rm th}=e^{- \beta H}/{\rm Tr}e^{- \beta H}$.
The gate fidelity takes values between $1/4$ and $1$, 
and the minimum gate fidelity $1/4$ implies that the output state is the completely mixed state and the gate operations fail.
In the case of fCH and iCH, by applying the unitary transformation $U_{\mathcal{T}}$,
the gate fidelity $F(l)$ is expressed in terms of the many-body correlation functions of the Ising model.
The results for the identity gates are shown in Fig.~\ref{Fig:Gate} for fCH and iCH (for the detailed calculations and the gate fidelities of the Hadamard gate derived analytically, see Appendix A and B).

For fCH, the gate fidelities exponentially decrease with increase of temperature to $1/4$ for any distance $l$.
For iCH, the gate fidelities change differently below/above the critical temperature,
namely, the gate fidelities also exhibit a transition at the critical temperature.
Above the critical temperature, the gate fidelities are close to $1/4$.
In contrast, the symmetry braking thermal state appearing below the
critical temperature leads dramatic improvement of the gate fidelities by exploiting its long-range order.
The temperature required to perform the gate operation reliably for iCH is much higher than that for fCH,
e.g., the gate fidelity for $l=8$ is almost $1$ if $T \lesssim 1.0 J$ for iCH and $T \lesssim 0.3J$ for fCH.
However, even for iCH, the fidelities just below the critical temperature is not large enough to reliably carry out
the gate operations, which motivates us to consider the topologically protected MBQC on the symmetry breaking thermal states on a 3D lattice.

{\it Topologically protected MBQC.---}
Topologically protected MBQC can be performed
with the cluster state on the RHG lattice \cite{RaussendorfA,RaussendorfB,RaussendorfC}.
The RHG lattice is defined by the set of cubes $Q$,
the set of faces $F_q$ on each cube $q \in Q$, and the set of edges
$E_f$ on each face $f \in F_q$.
The qubits are located on each face and edge
constituting the 3D cluster state,
where the nearest-neighbor stabilizer generators 
interact with each other, as shown in Fig. \ref{fig1} (c).
Similarly to the original case \cite{RaussendorfA,RaussendorfB,RaussendorfC},
the 3D cluster state is subdivided into three 
regions, vacuum $V$, defect $D$, and singular qubits $S$ used for
the topological protection, performing Clifford gates, and performing arbitrary single qubit gate, respectively.
In the following,
we consider the threshold value for the topological protection in the vacuum region $V$,
since it solely determines a threshold of quantum computation \cite{RaussendorfA,RaussendorfB,RaussendorfC}.

In the vacuum region $V$,
all qubits are measured in the $X$ basis.
Since $\prod _{f \in F_q} K_f = \prod _{f \in F_q} X_f$,
the parity of the measurement outcomes
on the six face qubits on each unit cell has to be even
for the ideal cluster state.
Thus the $Z$ errors on the face qubits, say error chain $C$, are
detected at the boundary $\partial C$,
which is called the error syndrome of $C$,
since the error chain $C$ anticommutes with $\prod _{f \in F_q} X_f$ there.
This is also the case for
the $Z$ errors on the edge qubits, say error chain $\bar C$,
since the edge qubits are the face qubits on the dual lattice.
In the case of the fCH on the RHG lattice \cite{RBH},
the error chains $C$ and $\bar C$, denoted by $\mathcal{C} \equiv (C, \bar C)$
are not correlated and can be treated independently.
However, in the case of the iCH, 
the primal and dual error chains of $\mathcal{C}$ 
are strongly correlated and have to be treated simultaneously.

By using the error syndrome $\partial \mathcal{C}$,
we infer the actual locations of the errors.
To this end,
the probability of a hypothetical error chain $\mathcal{C}'$,
which has the same error syndrome $\partial \mathcal{C}$,
is calculated to be
\begin{eqnarray}
p(\mathcal{C}'| \partial \mathcal{C}') 
= \mathcal{N}^{-1} \exp \left[{  \beta ' J \sum _{\langle f  \bar f\rangle} u_f^{C'} u_{\bar f}^{\bar C'}} \right]  \Big | _{\partial \mathcal{C}'= \partial \mathcal{C}}
\;,
\label{eq1}
\end{eqnarray}
where $\mathcal{N}$ is the normalization factor.
We used the knowledge that the errors occur 
with a ferromagnetic Ising-type distribution, 
which is characterized by a parameter $\beta '$
independently of the physical inverse temperature $\beta$.
The indicator function $u_f^{C'}$ is defined as $u_f^{C'} =-1$ 
for $f \in C'$ and $u_f^{C'} =1$  for $f \notin C'$,
specifying the location of the errors.
Since $\partial{C}' = \partial {C}$, we have
$\mathcal{C}'= \mathcal{C} + \mathcal{L}$ for
trivial loops (cycles) $\mathcal{L} \equiv (L,\bar{L})$, where $L$ is a trivial loop for the lattice 
and $\bar{L}$ is for the dual lattice, such that $\partial \mathcal{L}=0$.
In order to solve the loop condition,
we introduce gauge variables on the edges of primal and dual lattices
defined by $ P_f \equiv u_f^{L} = \prod _{e \in E_f} \sigma _e $
and  $\bar P _{\bar f} \equiv u_{\bar f}^{\bar L} = \prod _{\bar e \in \bar E_{\bar f}} \bar \sigma _ {\bar e} $.
In this parameterization, $\prod _{f \in F_q} P_{f} = 1$ and
$\partial L=0$
is automatically satisfied. As a result we obtain the Gibbs-Boltzman distribution
$p(\mathcal{C}'| \partial \mathcal{C})  = \mathcal{N}^{-1} e^{ - \beta'  H_\mathcal{C} ( \sigma ,\bar \sigma)}$
under a Hamiltonian given by
$
 H_{\mathcal{C}} ( \sigma , \bar \sigma ) = - J
 \sum _{\langle f\bar f \rangle } u_f^{C}  u_{\bar f}^{\bar C}P_f \bar P _{\bar f},
$
which we call correlated RPGM (cRPGM).
The sign of the two plaquette interaction $u_f^{C}  u_{\bar f}^{\bar C}$
representing the randomness of the model
is determined 
by the actual error chain $\mathcal{C}$
with the distribution 
$p(\mathcal{C})\equiv \mathcal{N}^{-1}e^{ \beta J \sum _{\langle f\bar f \rangle} u_f^{C} u_{\bar f} ^{\bar C}}$
parameterized by the physical inverse temperature $\beta$.

Since the threshold value for topologically protected 
MBQC corresponds to the critical point of the cRPGM \cite{Dennis},
our goal is to identify it.
Let us consider the optimal case of $\beta ' =\beta$ where
the actual and hypothetical error distributions are the same. 
This condition is referred to as the Nishimori line \cite{Nishimori} in spin glass theory.
In this case, 
the internal energy is given by
\begin{eqnarray*}
[\langle H_{\mathcal{C}}(\sigma , \bar \sigma ) \rangle _{\rm th}]_{\mathcal{C}}
&=& \sum _{\mathcal{C}} 
p(\mathcal{C})
\sum _{ \{ \sigma _e, \bar \sigma _{\bar e}\}} 
\frac{H_{\mathcal{C}}(\sigma , \bar \sigma) e^{ - \beta H_{\mathcal{C}}(\sigma , \bar \sigma )}}{\mathcal{Z}_{\mathcal{C}}	(\beta )},
\end{eqnarray*}
where $\langle \cdot \rangle_{\rm th}$ denotes the thermal average and
 $\mathcal{Z}_{\mathcal{C}}(\beta ) = \sum _{\{ \sigma _e, \bar \sigma _{\bar e} \}} e^{ - \beta H_{\mathcal{C}}(\sigma , \bar \sigma )}$ is the partition function.
We take the ensemble average of the error distributions $[\cdot ]_{\mathcal{C}}$ since the 
cRPGM has self-averaging property \cite{comment4}.
With the aid of the gauge symmetry \cite{Nishimori}, the Hamiltonian $H_\mathcal{C}(\sigma , \bar \sigma )$ is invariant 
under the following gauge transformations,
$u_f^{C}\rightarrow u_f^{C}  P'_f,
\;\;
 \sigma _{ e}\rightarrow \sigma _{e} \sigma' _{e},
\;\;
u_{\bar f}^{\bar C}\rightarrow u_{\bar f}^{\bar C}  \bar P'_{\bar f},
\;\;
 \bar \sigma _{\bar e}\rightarrow \bar \sigma _{\bar e} \bar \sigma' _{\bar e}$,
where $P'_{f}= \prod _{e\in E_f} \sigma ' _{e}$
and $\bar P'_{\bar f}= \prod _{\bar e\in \bar E_{\bar f}} \bar \sigma ' _{\bar e}$.
On the other hand,
these transformations changes the distribution $p(\mathcal{C})$
into
$\mathcal{N}^{-1}e^{ \beta J \sum _{\langle f, \bar f \rangle} u_f^{C} u_{\bar f} ^{\bar C} P_f '\bar P'_{\bar f}}
=\mathcal{N}^{-1}e^{- \beta  H_{\mathcal{C}}(\sigma' , \bar \sigma ')} \equiv p'(\mathcal{C})$,
which corresponds to the Gibbs-Boltzman distribution
for the cRPGM.
Since $\sum _{\{\sigma' , \bar \sigma' \}} p'(\mathcal{C})=\mathcal{Z}_{\mathcal{C}}(\beta)$, 
we obtain
\begin{eqnarray*}
&& [\langle H_{\mathcal{C}} (\sigma , \bar \sigma ) \rangle _{\rm th}]_{\mathcal{C}}
\\
&=&\frac{1}{\mathcal{N}}
\sum _{\mathcal{C}} 
\frac{1}{|\mathcal{L}|}\sum _{\{\sigma ' _e , \bar \sigma ' _{\bar e} \}}
 p'(\mathcal{C})
\sum _{ \{ \sigma _e, \bar \sigma _{\bar e}\}} \frac{H_{\mathcal{C}}(\sigma, \bar \sigma)
e^{ - \beta H_{\mathcal{C}}(\sigma, \bar \sigma)} }{\mathcal{Z}_{\mathcal{C}}(\beta )}
\\ 
&=&
\frac{1}{\mathcal{N}|\mathcal{L}|}
\sum _{ \{ \sigma _e, \bar \sigma _{\bar e}\}} \sum _{\mathcal{C}} 
H_{\mathcal{C}}(\sigma, \bar \sigma ) e^{ - \beta H_{\mathcal{C}} (\sigma ,\bar \sigma)}
\\
&=&
\mathcal{N} ^{-1}
\sum _{\mathcal{C}} 
  H_{\rm Ising} e^{ - \beta  H_{\rm Ising}}
 = \langle  H_{\rm Ising} \rangle _{\rm th},
\end{eqnarray*}
where $|\mathcal{L}|$ is the number of the 
loop configurations, 
$ H _{\rm Ising}= - J
 \sum _{\langle f,\bar f \rangle } u_f^{C}  u_{\bar f}^{\bar C}$,
and $\mathcal{N}$ is defined in Eq. (\ref{eq1}) as the partition function of the 
Ising model. 
For the transformation
from the third to forth lines,
we take the summation 
$\sum _{\sigma _e, \bar \sigma _{\bar e}} = |\mathcal{L}|$ 
by using the fact that
$u_f^{C}P_f = u_f^{C+L}$ and $u_{\bar{f}}^{\bar{C}} \bar{P}_{\bar{f}} = u_{\bar{f}}^{\bar{C}+\bar{L}}$
with trivial loops $\mathcal{L}$ and $\sum _{\mathcal {C}}=\sum _{ \mathcal{C}+\mathcal{L}}$.
Thus, the internal energy of cRPGM is equivalent to that of the Ising model without any randomness.

In the Ising model on the RHG lattice,
the internal energy has a non-analytical point at $T_c = 2.8$,
which is evaluated by the exchange Monte Carlo simulation \cite{HukushimaNemoto}.
Therefore we can conclude that
the internal energy of the cRPGM along the Nishimori line 
also has a non-analytical point at $T_c=2.8$,
which is the phase boundary of the Higgs (ordered) and confinement (disordered) phases 
\cite{Dennis,Wang,Ohno}.
In the Higgs phase,
the loop configurations $\mathcal{L}$ of large perimeters
are exponentially suppressed. Thus 
the logical error probability, which is characterized by
the loop configurations of non-trivial topology,
is decreased exponentially by increasing the size of the system
(see Appendix C for the decoding methods).
That is, the transition point of the performance in topologically protected MBQC
on the symmetry breaking thermal states
exactly determined by the critical temperature of the phase transition 
in the underlying physical system. 

The cluster state on the RHG lattice can
be also obtained from other lattices such as
simple cubic (SC), face-centered cubic (FCC), close-pack hexagonal (CPH) lattices
by measuring appropriate qubits in the $Z$ basis
as shown in Fig. \ref{fig1} (d)
in case for the SC lattice.
Since the thermal errors commute with the $Z$ basis measurements in our model, 
they do not induce any additional errors.
Also in these cases, by using the gauge transformation,
we can show that the thresholds for topologically protected MBQC
again given by the critical temperatures of the iCHs on those lattices.
The critical temperatures of the Ising models on the SC, FCC, and CPH lattices 
have been calculated numerically as $T_c=4.5, 9.3$, and $9.8$, respectively
\cite{Magalhaes},
which are higher than $T_c=2.8$ for the RHG lattice,
since each site interacts with more neighboring sites.
In comparison with $T_c = 0.59$ for the fCH,
the iCHs with the long-range order
relaxes the temperature required for topologically protected 
MBQC by more than one order of magnitude.
In the fCHs, 
the lattice structures do not change the threshold value for the topological protection
since the thermal errors occur independently for each qubit.
Contrary in the iCHs, 
the underlying lattice structures take a very important role
in robustness against the thermal excitation by making use of physical cooperative phenomena.

{\it Conclusions and Discussions.---}
We have demonstrated 
that physical cooperative phenomena of a system can
help MBQC on the system even at finite temperature. 
We have first shown that, in a square lattice, 
the gate fidelities of the identity gates for iCH are drastically improved compared to those for fCH below the critical temperature.
It has been also shown that the fidelities are not sufficiently large for performing MBQC reliably at the temperature just below the critical temperature even for iCH.
In the 3D cases,
MBQC on the thermal states are topologically protected
below the critical temperatures of the underlying physical system,
which allows us to perform MBQC on the symmetry breaking thermal states
even at much higher temperatures than the models without 
physical cooperative phenomena. 
A promising way to design these many-body interactions used in both fCH and iCH
is the stabilizer pumping scheme \cite{Weimer,Barreiro,Lanyon,Muller}.
Although achieving larger many-body interactions
requires more unitary operations in the scheme,
the required temperatures for performing topologically protected MBQC
is significantly relaxed for iCH.

In the present work,
we have considered only the Ising-type interaction 
in the stabilizer Hamiltonian.
We can also construct the iCHs,
which are unitarily equivalent to other spin models
such as the Potts, XY, and the Heisenberg models.
It is an interesting future work to study the relation between
the ordered phase and quantum information tasks in such models.
This will open up 
a new approach to make use of physical cooperative phenomena
for quantum information processing.

\begin{acknowledgments}
The authors  thank the fruitful discussions with H. Nishimori and S. Tanaka.
This work was supported by MEXT, Japan
(Project for Developing Innovation Systems, 
Grant-in-Aid for Scientific Research on Innovative Areas No. 20104003,
and Grant-in-Aid for
Young Scientists (B) No. 24740263),
and by JSPS (Grant No. 222812, Grant No. 23540463, and No. 23240001).

\end{acknowledgments}

\appendix

\section{Correlation functions in the interacting cluster Hamiltonians} \label{1}

We show the detailed
calculation of correlation functions in the interacting cluster Hamiltonians (iCHs) on a square lattice.
The Hamiltonian is given by
\begin{equation}
H_{\rm ic} = - J \sum_{\langle ij \rangle} K_i K_j,
\end{equation}
where the summation runs over all nearest neighbor bonds, $\langle ij \rangle$, of the square lattice $\mathcal{T}$. 
The thermal state of the iCH at temperature $T$
is denoted by $\rho_{\rm th} \equiv e^{ - \beta H_{\rm ic}}/ \mathcal{Z}_{ic}$,
where $\beta:=1/T$ is the inverse temperature, and 
$\mathcal{Z}_{ic} \equiv {\rm Tr} e^{ - \beta H_{\rm ic}}$ is the partition function.

The $k$-body correlation functions at temperature $T$ is defined by
\begin{equation}
\langle \prod_{n=1}^k K_{(i_n, j_n)} \rangle := {\rm Tr} \prod_{n=1}^k K_{(i_n, j_n)} \cdots K_{(i_k, j_k)} \rho_{\rm th},
\end{equation}
where $(i_n, j_n)$ denotes a coordinate of the square lattice. For simplicity, we assume $i_n<i_m$ and $j_n<j_m$ for $n<m$.

By the unitary transformation $U_{\mathcal{T}}$, which is a product of controlled-Z gates on all bonds,
iCH is transformed into the same form as the Hamiltonian of the Ising model,
$U_{\mathcal{T}} H_{\rm ic} U_{\mathcal{T}}^{\dagger} = -J\sum_{\langle ij \rangle} X_i X_j \equiv H_{\rm Ising}$.
Hence, the correlation functions are expressed in terms of those of the Ising model:
\begin{equation}
\langle \prod_{n=1}^k K_{(i_n, j_n)} \rangle = \langle \prod_{n=1}^k X_{(i_n, j_n)} \rangle_{\rm Ising},
\end{equation}
where $\langle A \rangle_{\rm Ising} \equiv \mathrm{Tr} A e^{ -\beta H_{\rm Ising}}/\mathcal{Z}_{\rm Ising}$.
In the following, we present the $k$-body correlation functions in the square-lattice Ising Hamiltonian for even and odd $k$.

\subsection{Even-body correlation functions} \label{SS:Even}
We present the exact formulas of 
even-body correlation functions on the $i$-th row, 
$\langle \prod_{n=1}^k X_{(i, j_n)} \rangle_{\rm Ising}$.
Since the Hamiltonian has a translational invariance, 
we consider only the correlation functions on the first row, $i=1$, without loss of generality
and we simply denote the correlation functions by $\langle \prod_{n=1}^{2k} X_{j_n} \rangle_{\rm Ising}$.
The correlation functions can be analytically obtained by using 
the method of a transfer matrix~\cite{MacCoy}, 
and the resultant exact formulae are given as follows.

We define a function 
\begin{equation}
c(\theta) = \frac{2z(1+z^2) - z^2(1-z^2) e^{i \theta} - (1-z^2)e^{- i \theta}}{\sqrt{ [(1+z^2)^2-2z(1-z^2) \cos \theta ]^2 - 4z^2 (1-z^2)^2 }},
\end{equation}
where $z:=\tanh \beta J$, and its Fourier transformation
\begin{equation}
C_r = \frac{1}{2\pi} \int_{- \pi}^{\pi} d\theta e^{-i r \theta} c(\theta).
\end{equation}
Using $C_r$, the correlation functions in the thermodynamic limit are analytically given by
\begin{equation}
\langle \prod_{n=1}^{2k} X_{j_n} \rangle_{\rm Ising}
=\sum_{\sigma} {\rm sign}(\sigma) \prod_{n=1}^k \prod_{l=j_{2n-1}}^{j_{2n}-1} C_{\sigma(l+1) - l -1}, \label{Eq:Even}
\end{equation}
where $\sigma$ is an element of a permutation group defined by
\begin{equation}
\sigma \in {\rm Perm} \bigcup_{n=1}^k \{j_{2n-1}+1,j_{2n-1}+2, \cdots, j_{2n}   \}.
\end{equation}

\subsection{Odd-body correlation functions}
In contrast to the even-body correlation functions,
the odd-body correlation functions are zero because of the symmetry of the Ising model.
To see this, apply a flip operator to all qubits, namely, a product of all $Z$ operators on each site.
Since the flip operator commutes with the Ising Hamiltonian $H_{\rm Ising}$ and anti-commutes with $X_{(i,j)}$ for any $(i,j)$,
we obtain
\begin{equation}
\langle \prod_{n=1}^{2k+1} X_{(i_n, j_n)}  \rangle_{\rm Ising} = -\langle \prod_{n=1}^{2k+1} X_{(i_n, j_n)} \rangle_{\rm Ising}, 
\end{equation}
leading to 
\begin{equation}
\langle \prod_{n=1}^{2k+1} X_{(i_n, j_n)} \rangle_{\rm Ising} = 0.
\end{equation}

Note that, when the symmetry of the system is broken,
the odd-body correlation functions are non-zero.
The odd-body correlation functions for symmetry breaking states have been analytically calculated under a certain assumption~\cite{HogeHoge}.
Here,
we numerically calculate them by the Monte-Carlo simulation without using the assumption since it has not been rigorously shown whether the assumption always holds, up to our knowledge.
Since the initial state of the Monte-Carlo simulation is breaking the symmetry,
it results in the correlation functions for the symmetry breaking states.

\subsection{Monte-Carlo simulation for correlation functions}

\begin{figure}
\begin{center}
\includegraphics[width=40mm]{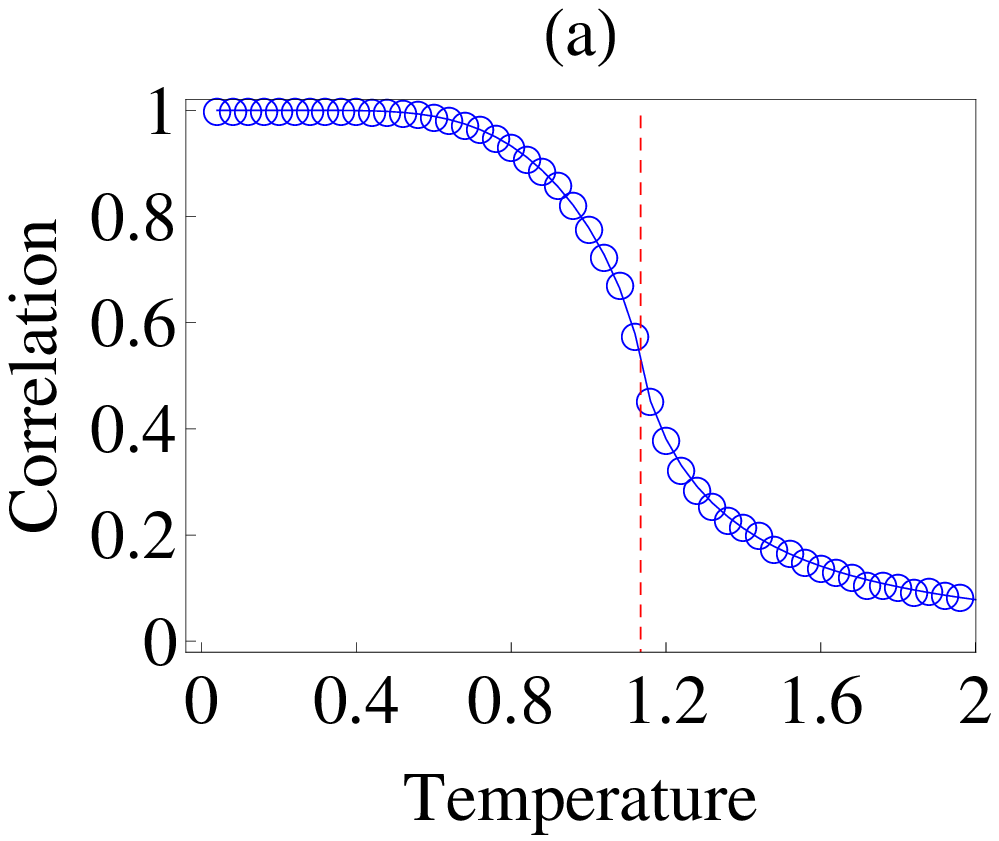}
\includegraphics[width=40mm]{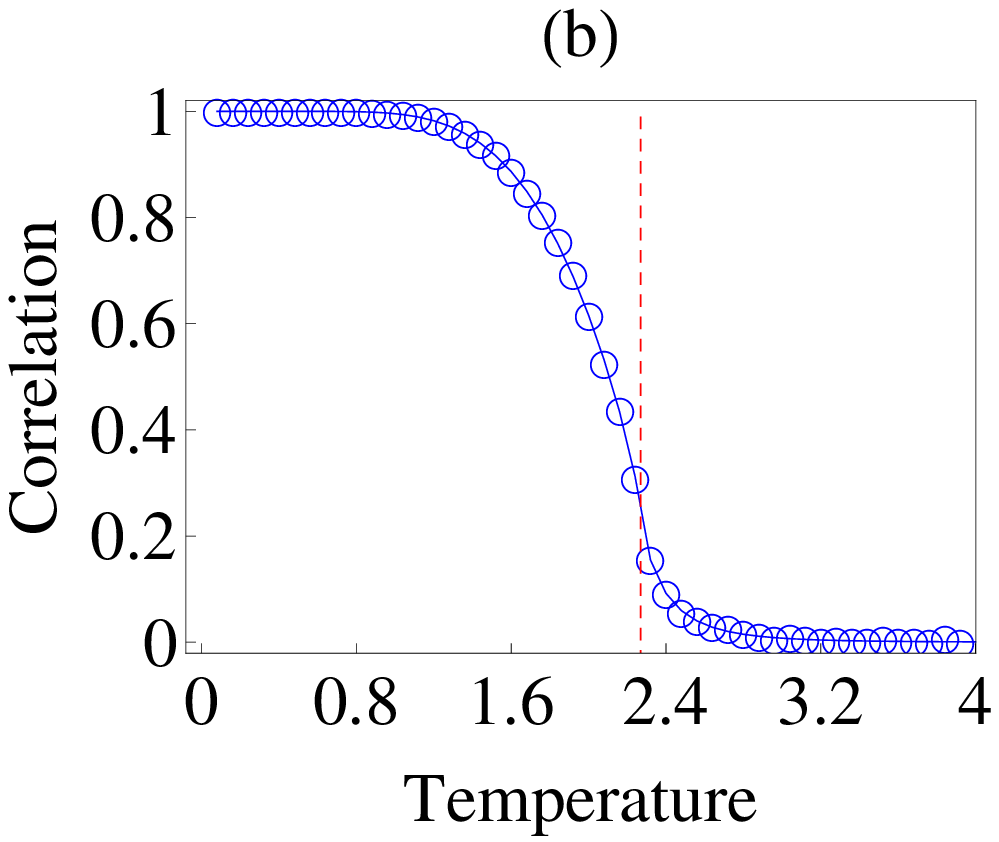}
\caption{The numerical results and the analytical results for (a) a correlation function $\langle X_{(1, 1)} X_{(1, 2)} X_{(1, 3)} X_{(1, 4)} \rangle_{\rm Ising}$ and (b) a correlation function $\langle \prod_{k=1}^6 X_{(1, 2k-1)} \rangle_{\rm Ising}$. The lines represent the exact results and the points ($\circ$) represent the numerical results.
In Fig. (b), the red vertical dashed line shows the critical temperature given by $T_c = 2/\ln[1+\sqrt{2}]$.
}
\label{Fig:Check}
\end{center}
\end{figure}

We provide numerical results of the odd-body correlation functions obtained by the Monte-Carlo simulation. 
The Monte-Carlo simulation performed with the lattice size $150 \times 150$ and the number of sampling $10^5$.
To check the finite size effects of the simulation,
we preliminarily compare the numerical results of the even-body correlation functions with the analytical results given in Sec.~\ref{SS:Even},
and confirm that the finite size effects are negligible in this case (see Fig.~\ref{Fig:Check}).

\begin{figure}
\begin{center}
\includegraphics[width=40mm]{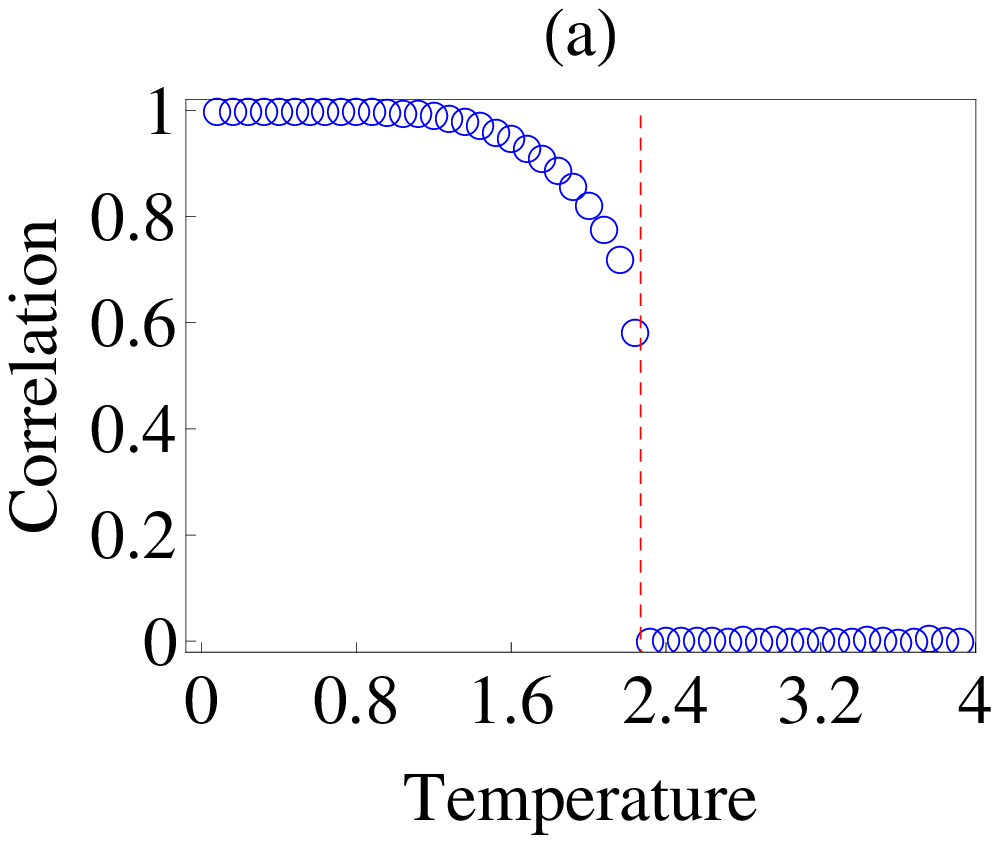}
\includegraphics[width=40mm]{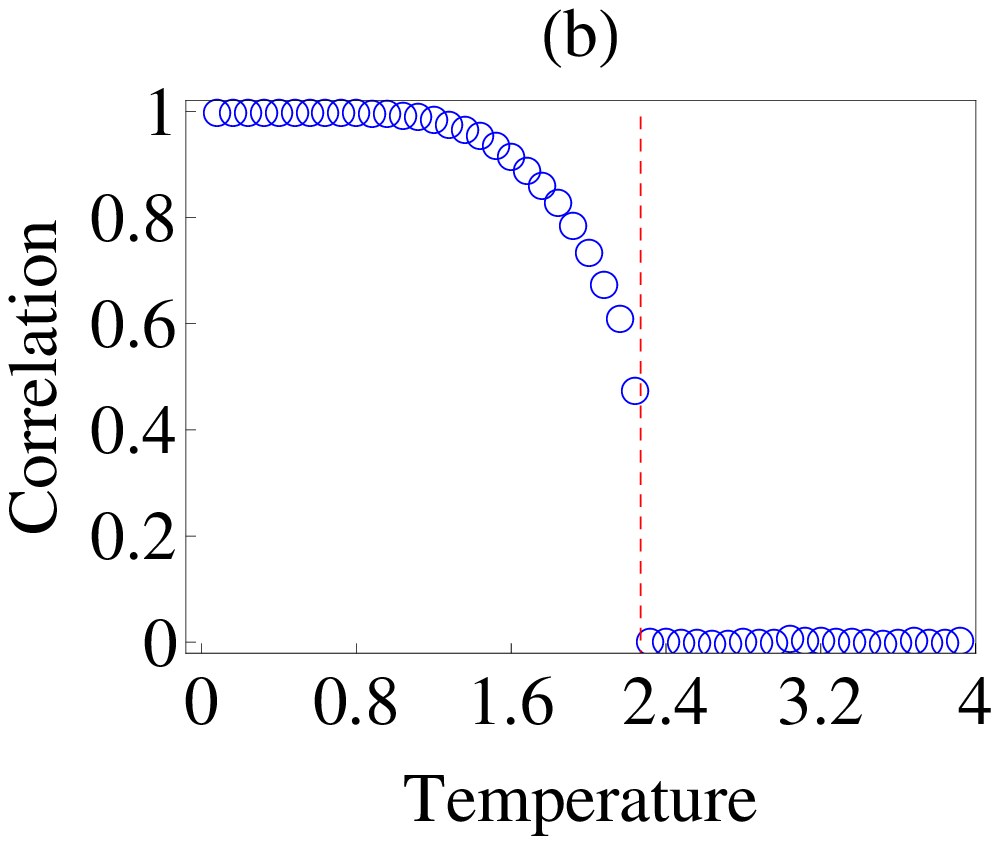}\\
\includegraphics[width=40mm]{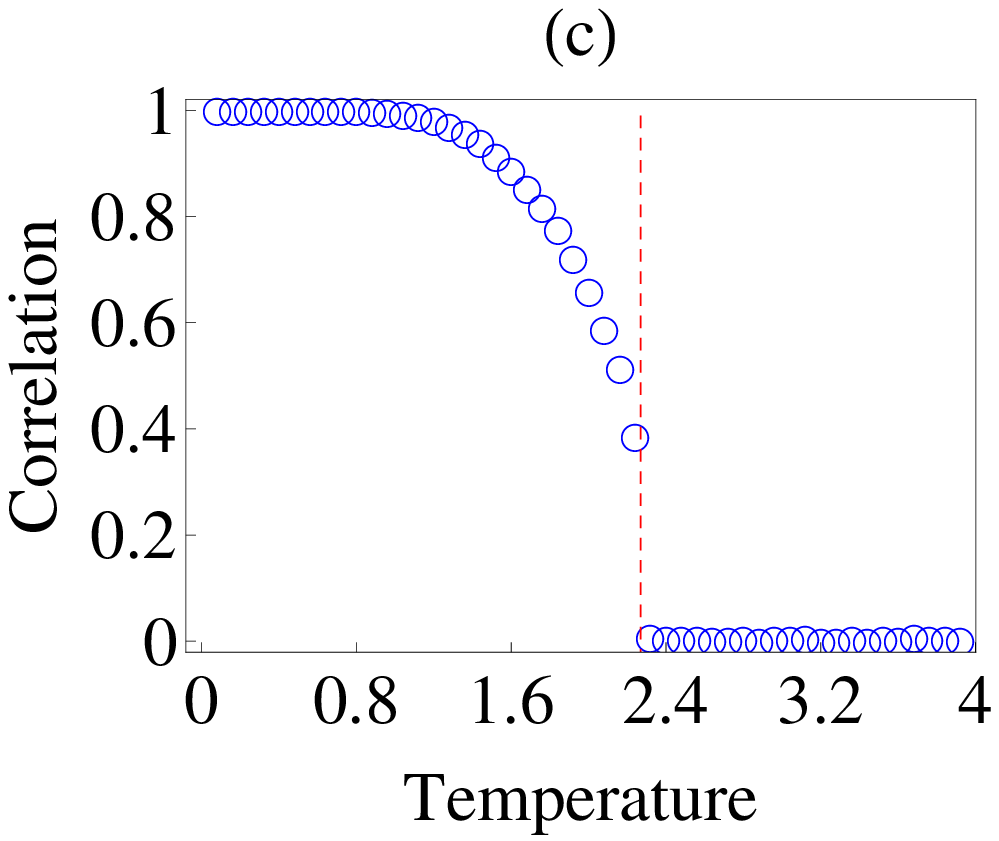}
\includegraphics[width=40mm]{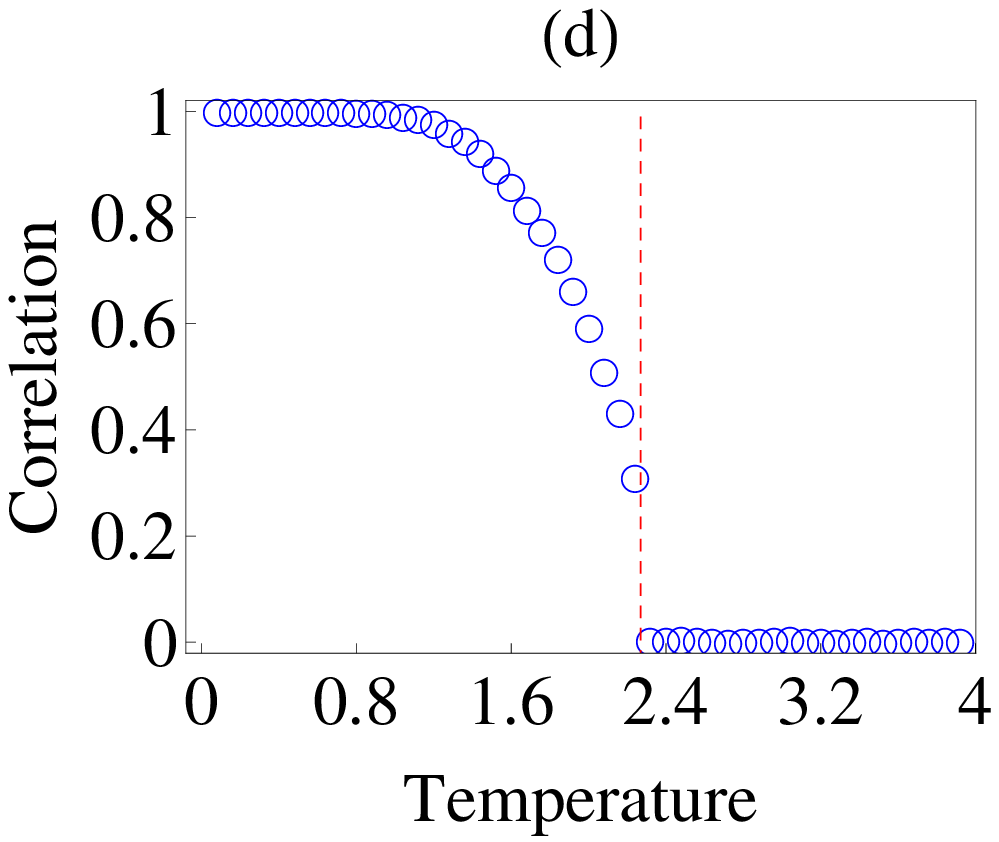}
\caption{The numerical results and the analytical results for the odd-body correlation functions $\langle \prod_{n=1}^{2k+1} X_{(1, n)} \rangle_{\rm Ising}$  for (a) $k=1$, (b) $k=2$, (c) $k=3$ and (d) $k=4$. In each figure, the red vertical dashed line shows the critical temperature $T_c = 2/\ln[1+\sqrt{2}]$.
}
\label{Fig:Odd}
\end{center}
\end{figure}

The numerical results for the odd-body correlation functions, $\langle \prod_{n=1}^{2k+1} X_{(1, n)} \rangle_{\rm Ising}$, are shown in Fig.~\ref{Fig:Odd}. The correlation functions are zero above the critical temperature. Below the critical temperature, the correlation functions are non-zero and are decreasing with increasing temperature. 
Note that, for larger $k$, the correlations decrease more quickly 
with increasing temperature, 
implying that the many-body correlations are not robust against the thermal excitations. The correlations at sufficiently low temperature are almost unity independently of $k$, that is, the spins are in an ordered direction.

\section{Gate fidelities of the Hadamard gate}
Based on the correlation functions, we calculate the gate fidelity of the Hadamard gate acting on the qubits at $(1,1)$ and $(1,2 l+1)$ where $l=1,2,\cdots$.
When a state is a thermal state $\rho_{\rm th}$, the gate fidelity is given by
\begin{equation*}
F(2 l+1) = {\rm Tr} \biggl[ \frac{I + \prod_{i=1}^{l} K_{(1,2i)} }{2} \frac{I + \prod_{i=1}^{l} K_{(1,2i)} }{2} \rho_{\rm th} \biggr].
\end{equation*}

\begin{figure}
\begin{center}
\includegraphics[width=65mm]{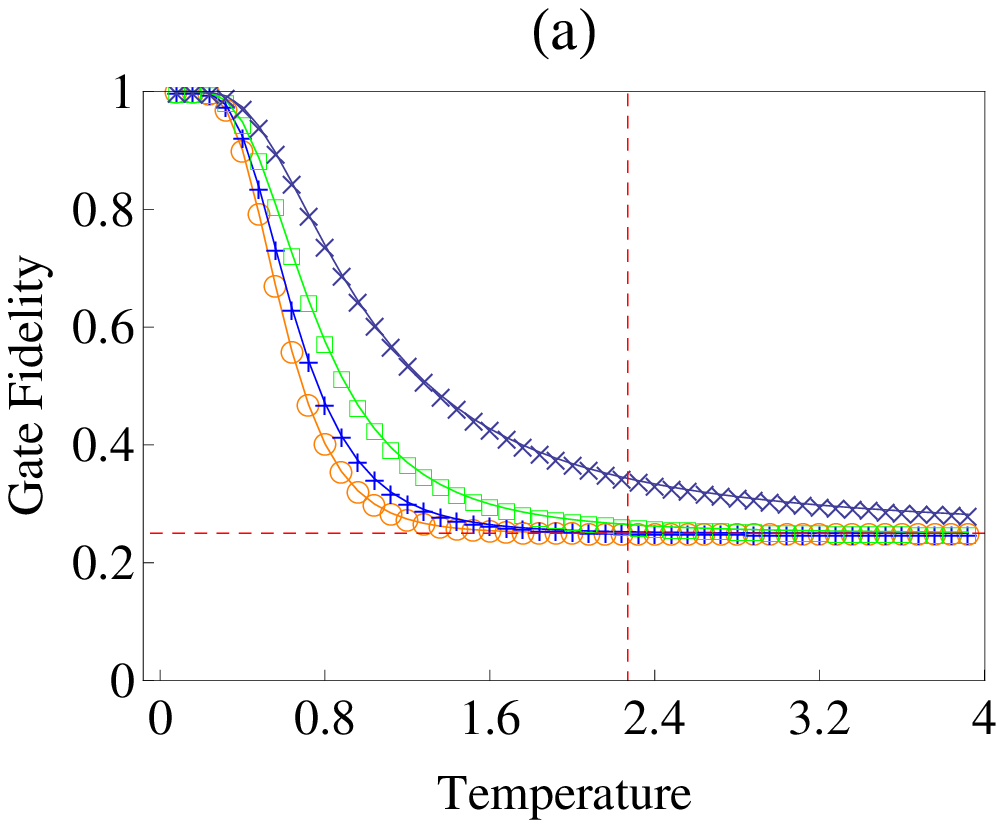}\\
\includegraphics[width=65mm]{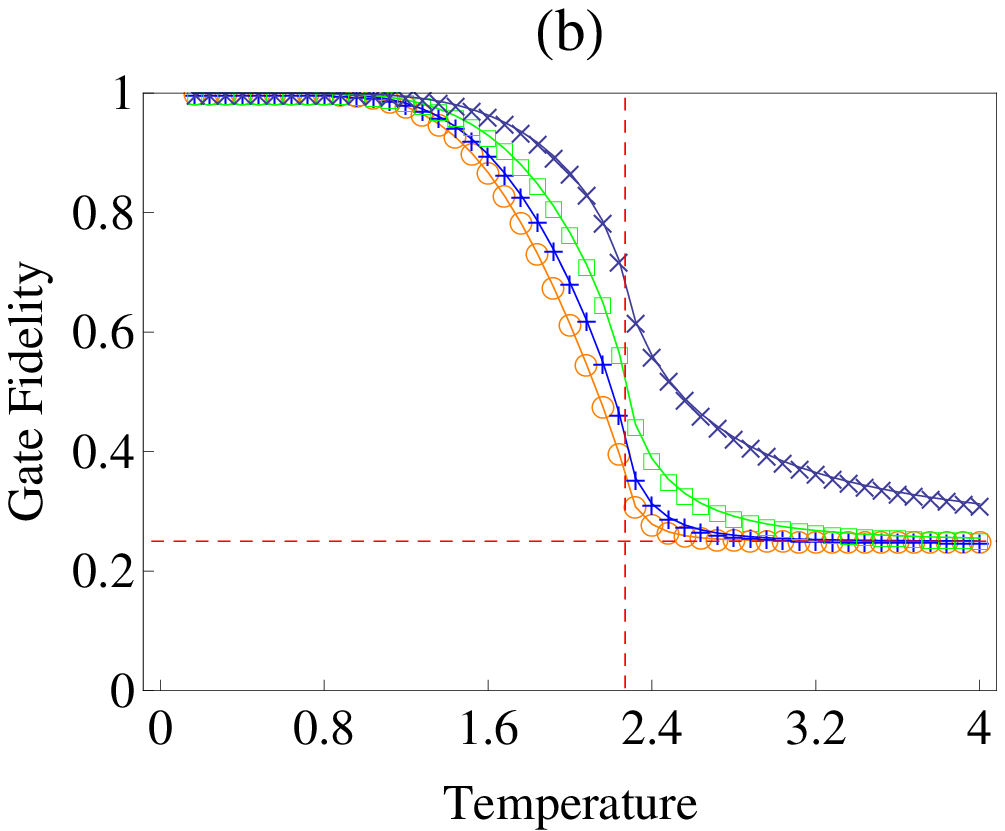}
\caption{The gate fidelity of the Hadamard gates for various distance $l$. Fig. (a) shows the gate fidelity for fCH,
$F (2 l+1)$ for $l=2$ ($\times$), $l=4$ ($\Box$), $l=6$ ($+$) and $l=8$ ($\circ$). 
Fig. (b) shows the gate fidelity for iCH,
$F(2l+1)$ for $l=2$ ($\times$), $l=4$ ($\Box$), $l=6$ ($+$) and $l=8$ ($\circ$). The vertical dashed line shows the critical temperature for the 2D iCH, $T_c/J= 2/\ln[1+\sqrt{2}]$ and the horizontal dashed line shows the minimum gate fidelity $F=1/4$.}
\label{Fig:HadGate}
\end{center}
\end{figure}

For a thermal state of a free cluster Hamiltonian (fCH), defined by $H_{\rm fc} = - J \sum_{i} K_i$, it is straightforward to calculate
$F(2 l +1)$ since thermal excitations on each qubit occur independently:
$F_{fc}(2 l +1) = (1+ \tanh^{l} \beta J)^2/4$.
For a thermal state of iCH, the gate fidelity can be calculated by using the results in Sec.~\ref{1}.
In particular, when $l$ is even, the gate fidelity $F(2 l+1)$ contains only even-body correlation functions
so that it can be analytically obtained by using Eq.~\eqref{Eq:Even}.
The results are shown in Fig.~\ref{Fig:HadGate}.

\begin{figure}
\begin{center}
\includegraphics[width=65mm]{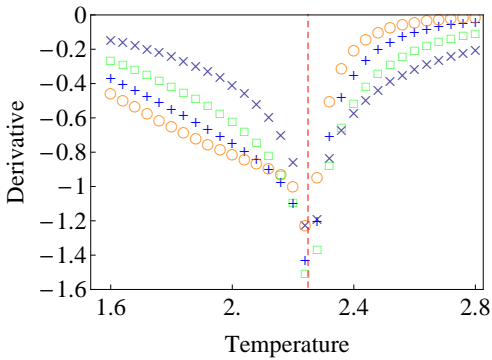}
\caption{The first derivative of the gate fidelity of the Hadamard gates for $l=2$ ($\times$), $l=4$ ($\Box$), $l=6$ ($+$) and $l=8$ ($\circ$). The vertical dashed line shows the critical temperature for the 2D iCH, $T_c/J= 2/\ln[1+\sqrt{2}]$.}
\label{Fig:HadDeri}
\end{center}
\end{figure}

Compared to the exponential decay for fCH, the gate fidelity for iCH changes its behavior at the critical temperature.
This transition of the gate fidelity is clearly observed by the derivative of the gate fidelity as shown in Fig.~\ref{Fig:HadDeri}. Similarly to the gate fidelity of the identity gate, the gate fidelity of the Hadamard gate for iCH is drastically improved below the critical temperature, although they are not sufficiently large  for performing MBQC reliably at the temperature just below the critical temperature.

\section{Decoding methods}
On the vacuum region $V$ of the RHG lattice,
we perform $X$ basis measurements for topological quantum error correction \cite{RaussendorfA,RaussendorfB,RaussendorfC}.
The thermal excitation results in the $Z$ errors on the face qubits on the primal and dual RHG lattices,
which are associated with the edges on the dual and primal lattices respectively,
and denoted by a chain $\mathcal{C}=(C,\bar C)$.
The $Z$ errors are detected at the boundary $\partial \mathcal{C}$ of the error chain $\mathcal{C}$,
since $\prod _{f \in F_{q}} K_f$ becomes odd there.
In the decoding procedure, 
we infer the location of the error chain $\mathcal{C}$
from the error syndrome $\partial \mathcal{C}$.

\begin{figure}
\begin{center}
\includegraphics[width=70mm]{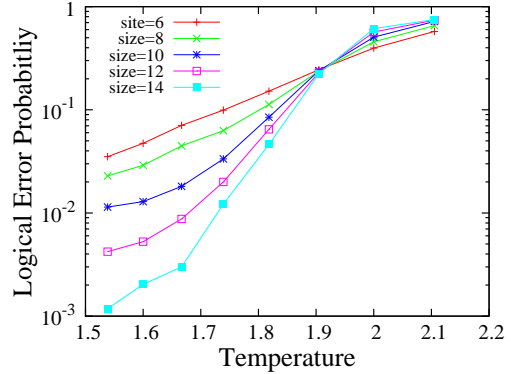}
\caption{
The logical error probability as a function of 
temperature, where the MWPMA is used for decoding.
The simulations are performed on the RHG lattices
of the size $N \times N \times N$ for $N=6,8,10,12,14$.
}
\label{Fig:MWPMA}
\end{center}
\end{figure}

In the case of fCH on the RHG lattice,
the $Z$ errors due to the thermal excitations
occur independently for each qubit.
Thus the minimum-weight-perfect-matching algorithm
(MWPMA) can be used as a suboptimal but good 
decoding method, resulting in 
the threshold value $T_c = 1/(\beta _{c} J)=0.57$ 
($p_{\beta J} = 2.9\%$) \cite{Wang},
which is close to the optimal one $T_c = 1/(\beta _{c} J)=0.59$ 
($p_{\beta J} = 3.3\%$) \cite{Ohno}.
In the case of iCH, the MWPM is far from optimal, since
the $Z$ errors are strongly correlated due to the interaction terms $K_i K_j$.
This is numerically observed in Fig.~\ref{Fig:MWPMA}.
The Fig. \ref{Fig:MWPMA} provides the logical error probability,
with which the error correction is failed, as a function of temperature $T=1/(\beta J)$.
If the temperature is smaller than $T=1.9$, the logical error probability exponentially decreases
with the lattice size $N$, which implies that the decoding succeeds.
However, this threshold $T=1.9$ is smaller than the optimal temperature $T_c = 2.8$ given by the 
critical temperature of the Ising model on the RHG lattice.

To achieve a successful decoding at the temperature close to the optimal temperature,
we should take into account the correlations between the $Z$ errors.
One way is making use of the free energy $\beta F_{\mathcal{C} '+ \mathcal{V} _{i}}(\beta) \equiv 
 -\ln  \mathcal{Z} _{\mathcal{C} ' + \mathcal{V}_{i}} ( \beta )$
of the cRPGM 
for hypothetical error chains $\mathcal{C} ' + \mathcal{V}_{i}$.
Here $\mathcal{C}'$ is an arbitrary error chain
that satisfies $\partial \mathcal{C}' = \partial \mathcal{C}$, 
and  $\mathcal{V}_{i}$ is a logical 
operator, which is represented by a non-trivial loop 
belonging to a homology class $i$ \cite{Dennis}.
In the decoding process, 
if the actual error chain $\mathcal{C}$ 
and estimated error chain $\mathcal{C}' + \mathcal{V}_{i}$
belong to the same homology class,
the error correction succeeds.
To succeed the error correction with a high probability,
we infer the most likely homology class from the error syndrome $\partial \mathcal{C}$.

The probability
that the error chains belong to the same homology class as that of $\mathcal {C}'+\mathcal{V}_{i}$ is given by   
\begin{eqnarray*}
p_{i} &=&
\frac{ \mathcal{Z}_{\mathcal {C}'+\mathcal{V}_{i}}(\beta) }
{\sum _{i} \mathcal{Z} _{\mathcal{C}'+\mathcal{V}_{i}}(\beta)}
\\
&=& \exp \bigl[
-\beta[ F_{\mathcal {C}'+\mathcal{L}_{i}}(\beta) - F_{\rm tot}(\beta)] \bigr],
 \end{eqnarray*}
 where $\beta F_{\rm tot}(\beta) \equiv - \ln \left[ \sum _{i} \mathcal{Z} _{\mathcal{C}'+\mathcal{V}_{i}}(\beta) \right]$
 is independent on the homology class $i$.
This indicates that the homology class $i _{\rm min}$
that has the minimum free energy 
is the most likely to occur. 
Thus if we perform the recovery operation
according to the estimated error chain $\mathcal{C}' + \mathcal{V}_{i_{\rm min}}$, the error correction succeeds with a high probability.
The logical error probability $p_{L}$ is the probability that the actual and estimated error chains
belong to different homology classes, namely, $p_{L}= 1 -  p_{i_{\rm min}}$.
In the Higgs (ordered) phase,
the logical error probability $p_{L}$ decreases
exponentially with the system size. 

The free energy can be calculated approximately
by numerical simulations, e.g., 
thermodynamic integration by use of data obtained in the Monte-Carlo simulations, and Population annealing~\cite{a} with the aid of the Jarzynski equality~\cite{b}.
For the decoding method using the free energy to be efficient,
the simulation time has to be polynomial in the system size. 
Although the relaxation time to equilibrium state takes exponentially long time
in spin-glass phases due to a large number of metastable configurations,
such spin-glass behaviors are not expected in the Higgs (ordered) phases.
Since we perform topologically protected MBQC in the Higgs phase appearing below the critical temperature,
it is expected that the error correction using the free energy
would be done efficiently \cite{Dennis}.

\end{document}